\documentclass[12pt,preprint]{aastex}

\newcommand{\be}{\begin{equation}}
\newcommand{\ee}{\end{equation}}

\shorttitle{$\omega$ Cen}
\shortauthors{T. Tsuchiya}

\begin{document}


\title{A Capture Scenario for Globular Cluster $\omega$ Centauri}


\author{Toshio Tsuchiya\altaffilmark{1},
Dana I. Dinescu\altaffilmark{2,3}, Vladimir I. Korchagin\altaffilmark{2,4}}

\altaffiltext{1}{Astronomishes Rechen-Institut,  M\"onchhofstra\ss e 12-14, 69120 Heidelberg, Germany, e-mail: tsuchiya@ari.uni-heidelberg.de, and
SGI Japan, Ltd. Kyoto city Skagura Bio-VIL 013, 668-3, 
Moto-Zaimokucho, Fushimi-ku, Kyoto, 612-8043 Japan}
\altaffiltext{2}{Astronomy Department, Yale University, P.O. Box 208101,
New Haven, CT 06520-8101, e-mail: dana@astro.yale.edu, vik@astro.yale.edu}
\altaffiltext{3}{Astronomical Institute of the Romanian Academy, Str.
Cutitul de Argint 5, RO-75212, Bucharest 28, Romania}
\altaffiltext{4}
{Institute of Physics, Stachki 194, Rostov-on-Don, Russia, 344090}

\begin{abstract}

We explore an accretion origin for $\omega$ Cen by
N-body modeling of the orbital decay and disruption of a 
Milky-Way dwarf satellite. This work is focused on
studying a particular satellite model that aims to reproduce the
present orbit of $\omega$ Cen, as recently determined from absolute 
proper motions. The model satellite is launched from 58 kpc from the
Galactic Center, on a radial, low-inclination orbit.
We find that a capture scenario can 
produce an  $\omega$ Cen-like object with the current 
low-energy orbit of the cluster. Our best model is
a nucleated dwarf galaxy with a Hernquist density profile that has
a mass of $8 \times 10^{9}$ M$_{\odot}$,
and a half-mass radius of 1.4 kpc. 

\end{abstract}


\section{Introduction}

Recent advances in understanding the nature of
$\omega$ Centauri, the most massive Milky Way globular cluster
($5 \times 10^6$ M$_{\odot}$, Meylan {\it et al.} 1995),
have now placed on a firmer ground its accretion origin
as opposed to formation within the Milky Way. Unlike
the majority of the Milky Way globular clusters, 
$\omega$ Cen is a complex chemical system with an 
extended star formation history. For instance, 
the multiple-peak metallicity distribution (e. g.,
Pancino {\it et al.} 2000, the 
correlation between age and metallicity (e.g., Hughes \& Wallerstein 2000), 
and the s-process heavy-element enhanced enrichment 
in cluster stars (e.g., Smith {\it et al.} 2000)
argue for a star formation history of 2 to 4 Gyr.
Furthermore, the correlation between the
enhancement of s-process elements with metallicity argues for a 
self-enrichment process within this system 
(Vanture, Wallerstein \& Suntzeff 2002)
as opposed to the build-up of the whole system by merging a few 
chemically distinct fragments. 
Gnedin {\it et al.} (2002) show that the present-day
$\omega$ Cen is not unique among globular clusters in its
ability to retain its own stellar ejecta, but its frequent passages 
through the Galactic disk should have swept out all of 
the s-process enriched gas from AGB stars. 
This is the case for all of the globular clusters within
10 kpc of the Galactic center, but not for $\omega$ Cen.

The internal kinematics of $\omega$ Cen lend further support to an
accretion origin.
The cluster rotates relatively rapidly causing
an ellipticity of $\sim 0.12$ (Merritt, Meylan, \& Mayor 1997),
while most of the Galactic globular clusters have an 
ellipticity of $\sim 0.03$ 
(Majewski {\it et al.} 2001). However, the metal rich stars 
([Fe/H] $\ge -1.2$) 
in $\omega$ Cen do not share the bulk rotation of the
metal poor stars, and have a lower velocity dispersion  
(Norris {\it et al.} 1997).  The metal rich stars
also show a North-South elongation in 
contradistinction to the  well-known East-West elongation of 
most of $\omega$ Cen's stars (Pancino {\it et al.} 2000).
The distinctive features of the metal-rich stars may be related to an
accretion event within $\omega$ Cen.

The extended star-formation history 
of $\omega$ Cen, the presence of stars formed from AGB-enriched gas,
and the evidence of a merger/accretion event point to a 
formation of $\omega$ Cen not
only away from the Milky Way, but also at the bottom 
of the potential well of a massive galaxy.
Thus, we wish to explore a previously established hypothesis:
$\omega$ Cen is the nucleus of a dwarf galaxy captured and 
disrupted by the Milky Way (Freeman 1993). 
A dwarf sinking to the center of the Galaxy is subject to dynamical
friction with simultaneous mass loss due to the interaction with
the Galactic tidal field. All of the dwarf is stripped off except for the
nucleus which, presumably, is left with the present-day orbit of $\omega$ Cen. 
This orbit is retrograde, of low energy and low inclination, and highly
eccentric. It has an
apocenter of  only 6.4 kpc, a pericenter of $\sim 1$ kpc, and a 
period of 120 Myr (Dinescu, Girard, \& van Altena 1999). 
Can the capture scenario produce the current orbit of $\omega$ Cen,
and if so, under what conditions? 
Zhao (2002) made the first attempt to model this accretion event using
a semi-analytical calculation.  
He found that, if the dwarf is
launched from 50 kpc from the Galactic center, the remnant 
will not decay to the present orbit. This is because strong tidal shocks
reduce the mass to the point that dynamical friction becomes
too inefficient to drag
the system to the current low-energy orbit of $\omega$ Cen.
Zhao's analysis was simplified by a number of assumptions. 
For instance, the mass outside the tidal radius is instantaneously stripped,
and the bulge and disk 
contributions to the dynamical drag were not taken into account
in the 50-kpc launch. 
We re-investigate the accretion scenario by means of
N-body simulations in order to follow the evolution of a dwarf galaxy 
falling into the Milky Way. 
%We use self-consistent models for the halo, 
%the bulge and the disk of the Milky Way to represent the
%Galaxy's potential. 
%We find that the accretion scenario can 
%produce an $\omega$ Cen-like object with the current orbit
%from a normal dwarf galaxy, and we present here the specific 
%characteristics of such a progenitor galaxy.

\section{The Models}

In our simulations, we use the model by Kuijken and Dubinski (1995) that
reproduces closely an exact solution of the set of collisionless 
Boltzmann and Poisson equations. 
The parameters of the model have been determined based on the observed
properties of the Milky Way. We assume the solar radius, 
the disk exponential scale length,
and the disk vertical scale height to be 
8 kpc, 3.5 kpc and 245 pc respectively.
The circular velocity of the disk at the solar radius 
is taken to be 220 km s$^{-1}$.
The surface density within 1.1 kpc of the disk plane in the model 
is equal to 69.8 M$_{\odot}$ pc$^{-2}$
with the disk contribution being 45.5 M$_{\odot}$pc$^{-2}$. 
%These values are close to the
%observed values for the surface density 
%71 $\pm$ 6 and 48 $\pm$ 9 M$_{\odot}$pc$^{-2}$
%respectively (Kuijken \& Gilmore 1991). 
With the generally accepted cutoff radius of the disk of 28 kpc, 
the disk's total mass in the model is 5 $\times$ 10$^{10}$ M$_{\odot}$.
The mass of the bulge is 0.75 $\times$ 10$^{10}$ M$_{\odot}$ within 2.4 kpc. 
The halo is approximated by a lowered isothermal sphere 
(Binney \& Tremaine 1987) with the central potential 
$\Psi(0)/\sigma_{0}^2 = $8. The halo mass inside 50 kpc is 
4.9 $\times$ 10$^{11}$ M$_{\odot}$
which is within the observationally inferred limits 5.4$^{+0.2}_{-3.6}$
(Wilkinson \& Evans 1999). A more detailed description of 
the Milky Way model can be found in Tsuchiya (2002).

We have examined the dynamics of the disrupting dwarf galaxy 
using two different density profiles:
1) a King profile, or lowered isothermal density distribution 
(King 1966), and 2) a Hernquist density profile
(Hernquist 1990). Our King profile has a deep central potential
($\Psi(0)/\sigma_{0}^{2} = 12$) with a 
density distribution in the central regions equivalent 
to an $r^{-2}$ cusp within our numerical resolution.
The highly-concentrated King profiles were chosen to model a 
satellite with a steep central density cusp in an attempt to 
mimic the nuclear region that would presumably end as an 
$\omega$ Cen-like object.
For the King profile, we have examined a set of
six models whose total masses and half-mass radii are summarized in Table 1
(K1 through K6).
The Hernquist model has a shallower density distribution than the King model,
with an $r^{-1}$ cusp in the central regions.
We have examined nine models of the Henquist 
profile whose parameters are also listed in Table 1 (H1 through H9).
In order to follow the dynamics of the central nucleus 
of a nucleated dwarf galaxy --- as the hypothetical progenitor of 
$\omega$ Cen --- we place at the center
of our preferred model dwarf galaxy
an extended 'particle' that has a mass  of
10$^7$ M$_{\odot}$, and $r_{1/2} = 35$ pc. We then follow the evolution 
of the dwarf galaxy and its central 'nucleus' until
the nucleus sinks to an orbit close to the center of the Milky Way. 
A more detailed description of the simulations and results is given elsewhere
(Tsuchia, Korchagin, \& Dinescu 2003).

%\section{N-Body Simulations}

The simulations have been done with the 
Self Consistent Field (SCF)-tree code that has a detailed description in 
Tsuchiya (2002).
A hierarchical tree algorithm has been used to calculate
the dynamics of the softened particles representing the disk,
the bulge of the Milky Way,
and the dynamics of the dwarf galaxy. 
We have used 70,000 particles to represent the disk, and
10,000 particles for the bulge.
The results presented here make use of a rigid halo, to avoid
a long computational time; these models did not take into account
dynamical friction on the halo.
We have however tested one model (i.e., our preferred
one, H4, see Table 1, and the following Section) using a ``living''
halo. This halo has been modeled with the SCF algorithm 
(Hernquist \& Ostriker 1992). 
The halo distribution is sampled by 100,000 particles.
We have taken the tolerance parameter  to be 0.7,
and the softening length to be 0.01 of the disk scale length 
which corresponds to 35 pc in our Milky Way model. 
We found that the decay time for the simulation with the living halo is
shorter than that for the rigid one (1.8 Gyr versus 
3 Gyr), but the relation between
the satellite's bound mass and apocenter distance is similar for the
two halos (Tsuchia, Korchagin, \& Dinescu 2003).
%This result indicates that the mass loss is governed by the
%orbit.

The King model dwarfs are sampled by 50,000 equal-mass particles.
For the Hernquist model dwarfs, we adopt a multi-mass N-body model.
This is because we need to achieve a better mass resolution in
the central regions of the dwarf galaxy, as the Hernquist model 
undergoes heavy mass loss down to
$\sim 10^{7}$ M$_{\odot}$, while the King models stop losing mass at
$\sim 10^{8}$ M$_{\odot}$ (see the next Section).
The Hernquist distribution function has been 
divided into three energy groups.
%where the energy is:
%$E = {1 \over 2}v^2 + \Phi ({\bf r})$.
%of the dwarf particles 
The lowest energy group is sampled by 50,000 particles with the
mass of each particle being 2 $\times$ 10$^{-7}$ of the total mass 
of the dwarf. The middle and the highest energy groups are sampled by 
particles with masses 
2$\times$ 10$^{-6}$, and 2$\times$ 10$^{-5}$ of the total mass of the dwarf, 
respectively, and the number of particles
in each group is 45,000. 
The dwarf was hence sampled by 140,000 particles in total.
The multi-mass model allows us to achieve a much better resolution of the
dynamics of the central regions of the dwarf compared to that  
given by equal mass models.
The dynamics of the central 10$^7$ M$_{\odot}$ of 
the dwarf is therefore sampled by
approximately 50,000 particles.

\section{The Sinking Satellite}

To approximately match the current orbit of $\omega$ Cen,
we have chosen the initial position and velocity of the satellite to be
(X, Y, Z) = (50, 0, 30) kpc, and ($V_X, V_Y, V_Z$) = (0, -20, 0) km s$^{-1}$.
Here, X is positive away from the Galactic center, Y is positive toward
Galactic rotation, and Z is positive toward the north Galactic pole; the Sun is
at  (X, Y, Z) = (8, 0, 0).
Due to the small initial velocity, the dwarf falls into a very
eccentric orbit with a typical pericenter of $\sim 1$ kpc. At each
pericentric passage, the dwarf experiences strong shocks mainly from the
encounters with the bulge  which strip off the dwarf's outer layers. 
Simultaneously with the mass loss, the dwarf sinks into a smaller orbit.

The King and the Hernquist models have qualitatively different 
evolutionary sequences. In Figure 1 we show the bound mass (at apocenter)
as a function of the apocentric distance for the King profiles (Fig. 1 a) 
and for the Hernquist profiles (Fig 1 b). For the same total mass, 
the King models with higher central densities (Table 1) sink deeper.
Once the dwarf mass decreases to 
$\sim 10^8$ M$_{\odot}$, there is practically no more orbital
decay. These models lose mass approximately exponentially with time. 
The ever decreasing  mass-loss rate requires a time that is longer than
the age of the Galaxy 
in order to reach a remnant with a bound mass of $\sim 10^7$ M$_{\odot}$
(Tsuchiya, Korchagin \& Dinescu 2003).
For the Hernquist model dwarfs the evolutionary sequences (Fig. 1 b)
show that the orbit decay stops at a given radius that
depends on the central density (see Table 1).
The dwarf continues to lose significant mass as opposed to the King
models. In fact,
all the models but H1 have completely dissolved after a time less than
5 Gyr. From these tests, we have chosen as our best representation 
of the progenitor of $\omega$ Cen, model H4.
In Figure 2 we show the galactocentric distance of the center of mass
of the tidally bound mass of the disrupting dwarf as a function of
time for the model that includes the nucleus, and for the model that
doesn't (see Section 2). 
After approximately 2 Gyr from the beginning of the simulations, 
the model with the nucleus establishes itself
into a low-energy orbit with pericentric and apocentric 
distances of about 1~kpc and 6~kpc.
The time evolution of the galactocentric radius of the 
nucleated model illustrates that indeed the nucleus 
has near-stationary radial oscillations after 2.5 Gyr (Fig. 2). The
non-nucleated model is completely disrupted in 2 Gyr.
This is more clearly seen in Figure 3, where the mass of the 
gravitationally bound particles is shown as a function of 
time for the nucleated model, and the non-nucleated one.
After 3 Gyr from the beginning of the simulation,
the mass of the gravitationally bound particles surrounding the central nucleus
is about 2 $\times$ 10$^7$ M$_\odot$ as can be seen in Figure 3.
By 4 Gyr, further stripping decreases the mass
 of the gravitationally bound mass of the nucleated dwarf to the mass of 
the central nucleus. We speculate that
this is a likely evolutionary history of the progenitor of $\omega$ Cen.

In Figure 4 we show a snapshot of the distribution of the stripped 
material taken at 3 Gyr. The left panel shows the distribution 
in the Galactic plane, while the right panels shows the
distribution perpendicular to the Galactic plane.
The stripped particles  fill in a flattened, disk-like volume
with a radius of about 10 kpc, and a vertical scale height
of about 2.5 kpc. 
The volume density of the stripped matter in the solar neighborhood is 
low however. The estimated
volume density of the stripped matter at $R \sim $ 8 kpc is about 
2 $\times$ 10$^{-5}$ M$_\odot$ pc$^{-3}$.
Dinescu (2002) analyzed the Beers {\it et al.} (2000) kinematically unbiased
catalog of metal poor stars, and found an excess of RR Lyrae stars in the
retrograde sample when compared to the prograde sample. The 
metallicity range of the sample where this excess was found
is [Fe/H] = -2.0 to -1.5,  and the orbital eccentricities are larger than 
0.8 (see Fig. 3 in Dinescu 2002).
These properties are representative of $\omega$ Cen and its
debris characteristics.
Specifically, 30 RR Lyrae stars were found
in the retrograde sample, and 15
in the prograde one, amounting to an excess of 15 stars
within a volume of 175 kpc$^3$.
Cseresnjes (2001) estimated a population of $\sim 8400$ RR Lyrae stars
in the Sagittarius dwarf spheroidal galaxy (Sgr), assuming that 
the galaxy lost about 50$\%$ of its total mass. Assuming also that RR Lyrae 
stars trace the light, he estimated an integrated absolute V-magnitude of -14.7
for the whole Sgr system. Estimating the number of RR Lyrae stars
in a dwarf galaxy from the  integrated absolute V-magnitude, 
assuming a mass-to-light ratio M/L$_{V}$ of 7 (e. g., Prada \& Burkert 2002), 
and using the
density of debris in the solar neighborhood
estimated by our preferred model, we obtain 
a number of 65 RR Lyrae stars in the volume of 175 kpc$^3$. 
Keeping in mind that the Beers (2000) catalog is not complete,
the number of excess RR Lyrae found by Dinescu (2002) is 
in reasonable agreement with the expected number from 
$\omega$ Cen progenitor's debris.

\section{Summary}

We have shown that a dwarf satellite with an initial mass of 
$8 \times 10^{9}$ M$_{\odot}$, a Hernquist density
profile of $r_{1/2} = 1.4$ kpc, 
and a nucleus of $10^{7}$ M$_{\odot}$ can be dragged
to the current low-energy orbit of $\omega$ Cen. 
The remnant is a central nucleus with a mass within a factor of 2 to a few
of that of present-day $\omega$ Cen.  
We suggest two refinements to the model to fully reconcile the 
current mass of $\omega$ Cen with that of the remnant.
The first one is purely computational in nature:
one can explore results from  
higher-resolution simulations that allow a $10^{6}$ M$_{\odot}$ nucleus.
The second one involves a more sophisticated model: 
one can replace the nucleus with a 
collection of individual particles in order to follow the mass loss  
due to frequent disk passages, presumably down to the present-day mass of 
$\omega$ Cen. In fact, Leon, Meylan \& Combes (2000) show that
$\omega$ Cen has tidal tails related to its last passage through the disk,
and indicating mass loss.

This study lends support to the hypothesis that $\omega$ Cen 
is the core of a disrupted
satellite galaxy captured from the outer halo, and it underlines
that the dwarf needs specific initial conditions in order to reproduce
$\omega$ Cen's characteristics.
Our results do not exclude the scenario proposed by Zhao (2002)
where the satellite is captured from the edge of the Galactic disk. 
This latter scenario however, implies ram-pressure stripping by the 
Galactic disk, thus challenging the complex abundance pattern
that $\omega$ Cen possesses. The debris from the parent galaxy forms
a disk-like structure within 6 kpc of the Galactic center.
Kinematical surveys within 1-2 kpc of the Sun 
should be able to detect such a structure, as a retrograde feature
(Dinescu 2002).

We thank T. Girard for many useful suggestions concerning the calculations
and W. van Altena for his comments on the manuscript. 
This research was supported in part by the NSF under grant AST 0098687.
VIK acknowledges R. Spurzem for providing partial
support at ARI, Heidelberg through the DFG grant RUS 17/112/02.

%\clearpage

\newpage
\begin{table}
\begin{center}
\begin{tabular}{ccc}

\multicolumn{3}{c}{Table 1. Model Parameters} \\
\hline
\hline
\\
\multicolumn{1}{l}{Model} & \multicolumn{1}{c}{$M_{tot} (M_{\odot})$} & 
\multicolumn{1}{c}{$r_{1/2}$ (kpc)} \\
\hline 
\\
K1 & $4 \times 10^{9}$  & 1.0 \\
K2 & $4 \times 10^{9}$  & 2.0 \\
K3 & $4 \times 10^{9}$  & 4.0 \\
K4 & $8 \times 10^{9}$  & 1.0 \\
K5 & $8 \times 10^{9}$  & 2.0 \\
K6 & $8 \times 10^{9}$  & 4.0 \\ \\
H1 & $4 \times 10^{9}$  & 1.0 \\
H2 & $4 \times 10^{9}$  & 2.0 \\
H3 & $4 \times 10^{9}$  & 4.0 \\
H4 & $8 \times 10^{9}$  & 1.414 \\
H5 & $8 \times 10^{9}$  & 2.282 \\
H6 & $8 \times 10^{9}$  & 5.657 \\
H7 & $16 \times 10^{9}$  & 2.0 \\
H8 & $16 \times 10^{9}$  & 4.0 \\
H9 & $16 \times 10^{9}$  & 8.0 \\

\hline
\end{tabular}
\end{center}
\end{table}

\figcaption[f1.eps]{Bound mass versus apocentric distance diagram showing the evolution of a
sinking dwarf satellite. Panel a) shows the King model dwarfs, while panel b)
 the Hernquist model dwarfs.}

\figcaption[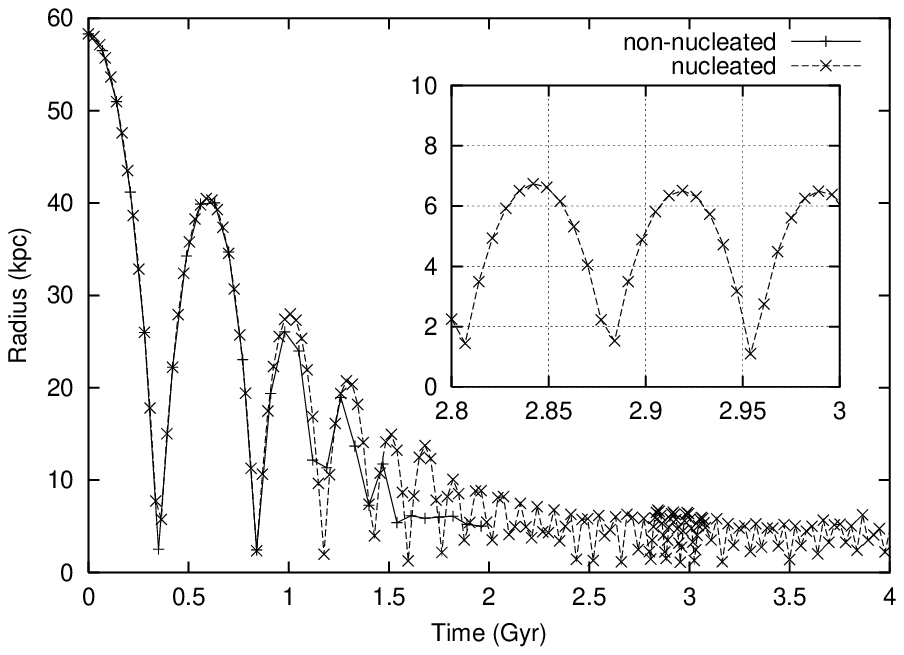]{The time dependence of the radius of a sinking dwarf 
satellite: nucleated - x symbols, non-nucleated - cross symbols.
The inset box shows the time dependence for the radius between 2.8 and 3 Gyr.}

\figcaption[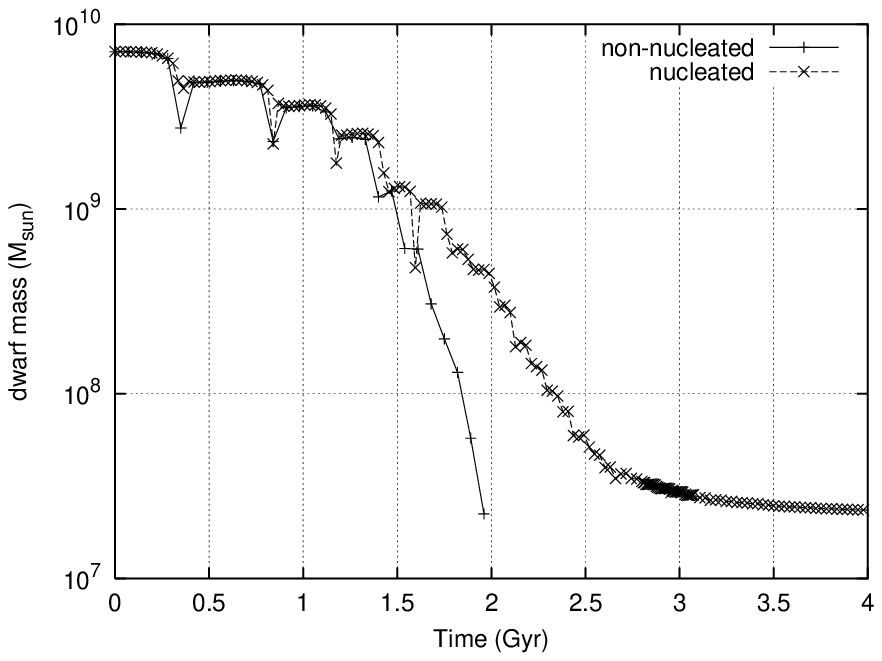]{The bound mass of the dwarf as a function of time. 
Symbols are the same as in Fig. 2. The non-nucleated dwarf is 
completely disrupted in 2 Gyr. 
The bound mass of the nucleated dwarf approaches a mass
of about 10$^7$ M$_{\odot}$ after 3 Gyr of evolution.}

\figcaption[f4.ps]{A snapshot of the stripped particles taken at 3 Gyr.
The left panel shows the distribution in the Galactic plane, while the
right panel, that perpendicular to the Galactic plane.}

%%%UCP%%%
\newpage
\plotone{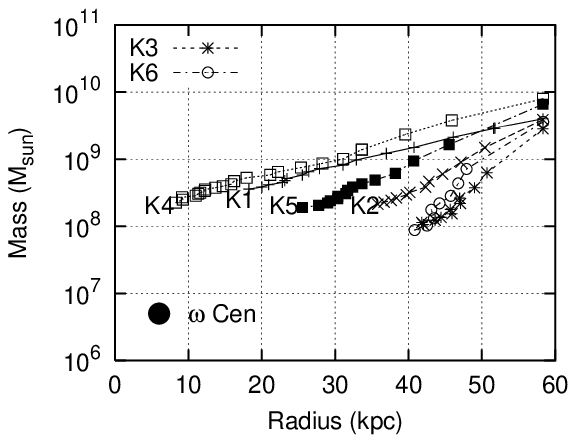}
\newpage
\plotone{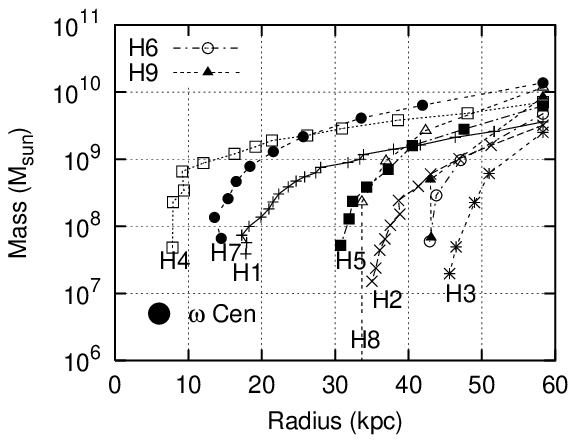}
\newpage
\plotone{f2.eps}
\newpage
\plotone{f3.eps}
\newpage
\plotone{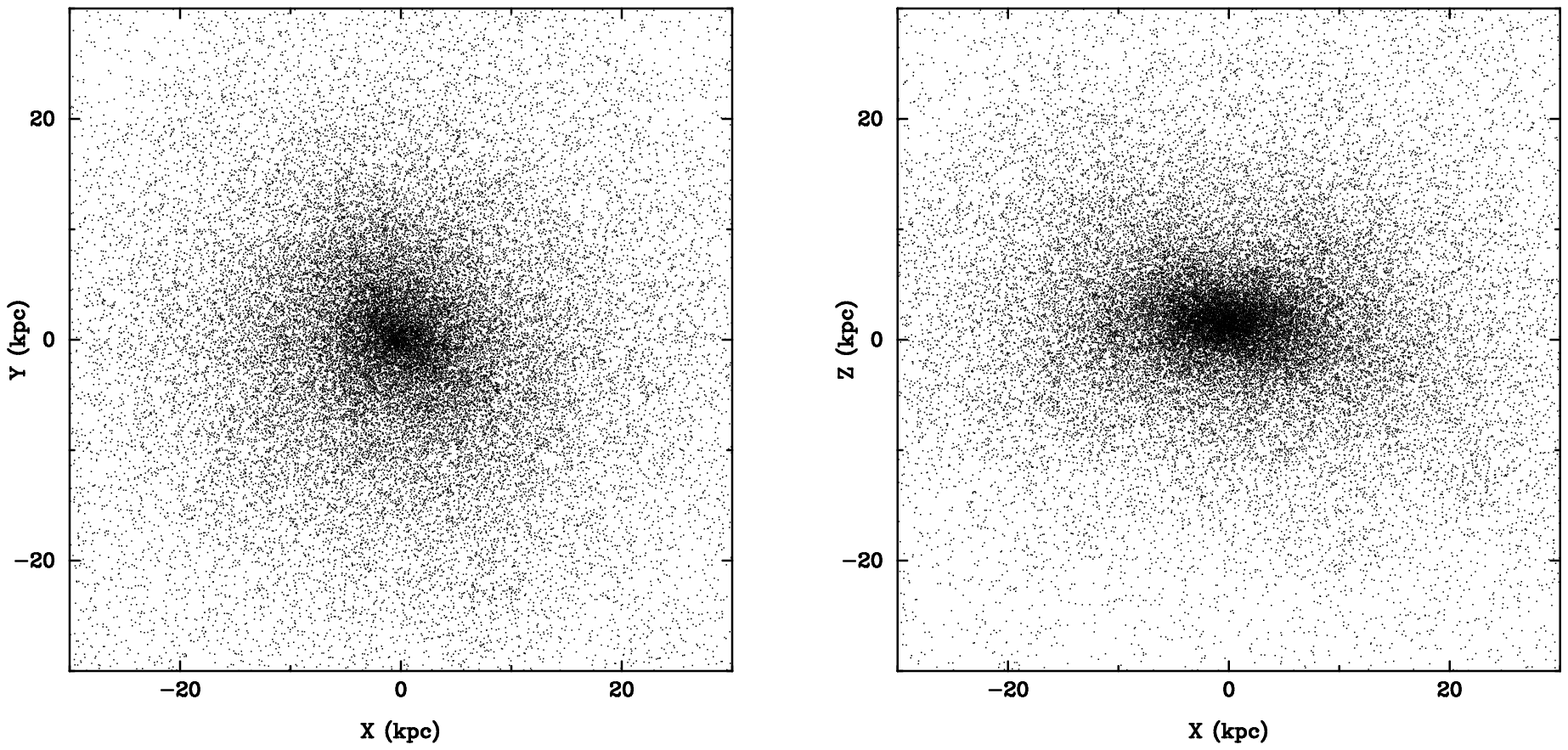}

\end{document}